\documentclass[twocolumn,nofootinbib,prl,aps,showpacs,superscriptaddress,floatfix]{revtex4-1}
\usepackage[latin1]{inputenc}
\usepackage{bm}
\usepackage[usenames]{color}
\usepackage{multirow}
\usepackage{amssymb}
\usepackage{amsbsy}
\usepackage{amsmath}
\usepackage{graphicx}
\usepackage{epsfig}
\usepackage{placeins}
\usepackage{ulem}

\makeatletter

\begin{document}

\title{Eigenstate Thermalization Hypothesis and Quantum Jarzynski Relation for Pure Initial States}

\author{F. Jin}
\email{f.jin@fz-juelich.de}
\affiliation{Institute for Advanced Simulation, J\"ulich Supercomputing Centre,
Forschungszentrum J\"ulich, D-52425 J\"ulich, Germany}

\author{R. Steinigeweg}
\email{rsteinig@uos.de}
\affiliation{Department of Physics, University of Osnabr\"uck, D-49069 Osnabr\"uck, Germany}

\author{H. De Raedt}
\affiliation{Zernike Institute for Advanced Materials, University of Groningen, NL-9747AG Groningen, The Netherlands}

\author{K. Michielsen}
\affiliation{Institute for Advanced Simulation, J\"ulich Supercomputing Centre, Forschungszentrum J\"ulich, D-52425 J\"ulich, Germany}
\affiliation{RWTH Aachen University, D-52056 Aachen, Germany}

\author{M. Campisi}
\email{michele.campisi@sns.it}
\affiliation{NEST, Scuola Normale Superiore \& Istituto Nanoscienze-CNR, I-56126 Pisa, Italy}

\author{J. Gemmer}
\email{jgemmer@uos.de}
\affiliation{Department of Physics, University of Osnabr\"uck, D-49069 Osnabr\"uck, Germany}

\begin{abstract} Since the first suggestion of the Jarzynski equality many
derivations of this equality have been presented in both, the classical and the
quantum context. While the approaches and settings greatly differ from one to
another, they all appear to rely on the initial state being a thermal Gibbs
state. Here, we present an investigation of work distributions in driven
isolated quantum systems, starting off from pure states that are close to
energy eigenstates of the initial Hamiltonian. We find that, for the
nonintegrable system in quest, the Jarzynski equality is fulfilled to good
accuracy. 
\end{abstract}



\maketitle

{\it Introduction.} The last decades have witnessed a renewed interest in the
old question if and how closed finite quantum systems approach thermal equilibrium.
Equilibration and thermalization have been theoretically discussed for both,
fairly abstract \cite{popescu2006, goldstein2006, reimann2008, reimann2015,
eisert2015, gogolin2015} and more specific systems of condensed-matter type
\cite{rigol2008, steinigeweg2013, beugeling2014, steinigeweg2014-1}. Key concepts
in this discussion are typicality (or concentration of measure) and the
eigenstate thermalization  hypothesis (ETH). With the advent of experiments on
ultracold atoms, some of the theoretical results have even become
testable. As of today, the mere existence of some sort of equilibrium in closed
quantum system has been the most widely addressed question. However, lately the
dynamical approach to equilibrium has been intensely investigated \cite{malabarba2014,
reimann2016}. Here, crucial questions are relaxation times but also the degree of
agreement of quantum dynamics with standard statistical relaxation concepts like
master or Fokker-Planck equations, stochastic processes, etc.\ \cite{niemeyer2014,
gemmer2014, schmidtke2016} The crucial feature that discriminates these types of
analysis from standard open-systems concepts, like quantum master equations, is
the fact that the statistical dynamics emerge from the systems themselves, i.e.,
are not induced by any bath.

Also fluctuation theorems have been and continue to be a central topic in the
field of statistical mechanics \cite{Seifert08EJPB64}. The Jarzynski relation (JR),
making general statements on work that has to be invested to drive processes
also and especially far from equilibrium, is a prime example of such a
fluctuation theorem. Many derivations of the JR from various starting grounds
have been presented. These include classical Hamiltonian dynamics, stochastic
dynamics such as Langevin or master equations, as well as quantum mechanical
starting points \cite{Seifert08EJPB64, Esposito09RMP81, Jarzynski11ARCMP2,
Roncaglia14PRL113, haenggi2015}. However, all these derivations assume that
the system that is acted on with some kind of ``force'' is strictly in a Gibbsian
equilibrium state before the process starts. This starting point differs from
the progresses in the field of thermalization: There, the general features of
thermodynamical relaxation are found to emerge entirely from the system itself
without the necessity of evoking external baths or specifying initial states
in detail. Clearly, the preparation of a strictly Gibbsian initial state
requires the coupling to a bath prior to starting the process.

\begin{figure}[t]
\includegraphics[width=0.80\columnwidth]{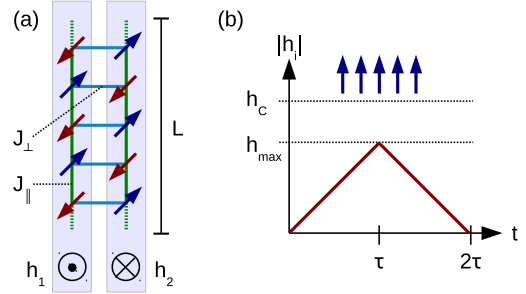}
\caption{Sketch of the (a) Heisenberg $S=1/2$ ladder studied and (b) the
time-dependent magnetic field applied.}
\label{fig1}
\end{figure}

In this Letter, we study the question whether or not the JR is valid
with a system starting in a state other than a Gibbs state. Since
counterexamples can be constructed, any affirmative answer cannot hold
for any quantum system and for any process protocol. In fact, previous
works \cite{Talkner08PRE77, Campisi08PRE78, talkner2013, campisi2011}
have shown that, when the initial state is microcanonical, the JR
does not follow, but a related entropy-from-work relation
emerges instead. The question remains, however,
if and under what conditions the JR holds approximately for non-canonical
initial states. Thus, the emphasis in the search for the origins of the
JR's validity is shifted from specifying the initial state to specifying
the nature of the system.

{\it JR and ETH.} To further clarify this, consider the standard
setup of the quantum JR for closed systems. It is based on a two-measurement
scheme: If the system is at energy $E_\text{ini}$ before the process, then
there is a conditional probability $T = T(E_\text{fin} | E_\text{ini},
\lambda (t))$ (with $\lambda(t)$ being the protocol) to find the system
at $E_\text{fin}$ after the process \cite{note}. Let $W = E_\text{fin}
- E_\text{ini}$ be the work associated with this transition. The average
of the exponentiated work $\langle e^{-\beta W}\rangle$ can now be written
as 
\begin{equation}
\langle e^{-\beta W}\rangle = \! \sum_{E_\text{fin},E_\text{ini}} \! T  \,
e^{-\beta ( E_\text{fin} -E_\text{ini})} P_\text{ini}(E_\text{ini}) \, .
\label{jarqua}
\end{equation}
Obviously, $\langle e^{-\beta W}\rangle$ depends on $P_\text{ini}(E_\text{ini})$.
It is well-known that the JR $\langle e^{-\beta W} \rangle = e^{-\beta \Delta F}$
always holds for initial Gibbs states $P_\text{ini}(E_\text{ini}) \propto e^{-\beta
E_\text{ini}}$, regardless of the system and the protocol. Much less is known
on other initial states, e.g., initial energy eigenstates $P_\text{ini}(E_\text{ini})
\approx \delta_{E_\text{ini}, E_n}$ with $E_n$ being the energy of an eigenstate.
These states are in center of our Letter.

It is very important to note that this question can be recast as a question on the
validity of the ETH in a specific sense: As shown in \cite{Talkner08PRE77}, the
average exponentiated work can be written as the expectation value
\begin{equation}
\langle e^{-\beta W}\rangle= \langle e^{-\beta H^H_\text{fin}} e^{\beta H_\text{ini}}
\rangle_{\text{diag}} \, , \label{wonob}
\end{equation}
where $H_\text{fin/ini}$ are the final/initial Hamiltonian (with the index $H$
indicating the Heisenberg picture) and $\langle \cdots \rangle_{\text{diag}}$
denotes the average over the diagonal part of the initial density matrix
w.r.t.\ the eigenbasis of $H_\text{ini}$ \cite{Talkner08PRE77}. Let 
$\langle \cdots \rangle_{\text{can}/ \text{mic}}$ denote averages over
canonical/microcanonical states, which are both diagonal in the above sense. While
the standard JR $\langle e^{-\beta W}\rangle = \langle e^{-\beta H^H_\text{fin}}
e^{-\beta H_\text{ini}} \rangle_{\text{can}} = e^{-\beta \Delta F}$ always holds,
our questions can be reformulated as
\begin{equation}
\langle e^{-\beta H^H_\text{fin}} e^{\beta H_\text{ini}}\rangle_{\text{mic}}
\stackrel{?}{=}\langle e^{-\beta H^H_\text{fin}} e^{\beta H_\text{ini}}
\rangle_{\text{can}} \, .
\label{ETH}
\end{equation}
The validity of this equation is claimed by the ETH (even though the operator in
the average is non-Hermitian). Since each protocol yields a different $H^H_\text{fin}$,
the JR's validity for microcanonical states is equivalent to the ETH's validity
for a set of different operators. So far, however, no general principle guarantees
the applicability of the ETH, except for large quantum systems with a direct classical
counterpart \cite{srednicki1994} or systems involving random matrices \cite{deutsch1991}.
While the ETH is expected to hold for non-integrable systems and few-body observables,  
$e^{-\beta H^H_\text{fin}} e^{\beta H_\text{ini}}$ is not such an operator. Thus,
investigating the JR's validity for microcanonical states is an highly non-trivial
endeavor.
 
In this Letter, we use numerical methods to prepare an energetically firmly
concentrated initial state and to propagate it according to the Schr\"odinger
equation for a complex spin system with a strongly time-dependent Hamiltonian. Due
to the initial state being sharp in energy, the eventual energy-probability
distribution is interpreted as a work-probability distribution and thus checked
for agreement with the JR, including a careful finite-size scaling.
As there is no thermal initial state, we use for the inverse temperature
$\beta$ in the JR the standard definition $\text{d} S / \text{d} E = \beta$ and
resort to the microcanonical entropy $S = \ln n(E)$, where $n(E)$ is the density
of energy eigenstates (DOS). Thus, $\beta$ depends on the spectrum. Since $\beta$
also depends on $E$, we evaluate $\beta$ at the initial energy $E_\text{ini}$.

{\it Spin model and time-dependent magnetic field.} The choice of our specific
spin model including all its parameters roots in its being a prime example for
the emergence of thermodynamical behavior in closed, small quantum systems. The
quantum dynamics of certain observables have been found to be in remarkable
accord with an irreversible Fokker-Plank equation for the undriven system and
with a Markovian stochastic process in a more detailed sense \cite{niemeyer2014}.

As shown in Fig.\ \ref{fig1}, we study an anisotropic spin-$1/2$ Heisenberg ladder
with the rung coupling being significantly weaker than the leg coupling. Specifically,
the Hamiltonian $H = J_\parallel H_\parallel + J_\perp H_\perp$ consists of a leg
part $H_\parallel$ and a rung part $H_\perp$,
\begin{eqnarray}
\label{ham}
H_\parallel \!\! &=& \!\! \sum_{i=1}^{L-1}\sum_{k=1}^2 S_{i,k}^x S_{i+1,k}^x
+  S_{i,k}^y S_{i+1,k}^y + \Delta S_{i,k}^z S_{i+1,k}^z \, , \nonumber \\
H_\perp \!\! &=&  \!\! \sum_{i=1}^{L} S_{i,1}^x S_{i+1,2}^x + S_{i,1}^y
S_{i+1,2}^y + \Delta S_{i,1}^z S_{i+1,2}^z \, ,
\end{eqnarray}
where $S_{i,k}^{x,y,z}$ are spin-$1/2$ operators at site $(i,k)$. $J_{\parallel,
\perp} > 0$ are antiferromagnetic exchange coupling constants with $J_\perp =
0.2 \, J_\parallel$, $\Delta = 0.6$ is the exchange anisotropy in the $z$
direction, and $L$ is the number of sites in each leg. We set $J_\parallel = 1$
throughout this work.

A magnetic field ($h$) is turned on once the time evolution starts. The field is
uniform along each individual leg, pointing in the positive $z$ direction on
one leg and in the negative $z$ direction on the other. This field is linearly
ramped up in time from zero to $h_{max}$ for a certain time $\tau$, and then linearly ramped
down for the same time with the same slope. Thus, the field starts of at zero
and ends at zero, i.e., initial and final Hamiltonian are identical. More
precisely, we model the field by
\begin{equation}
h(t) = -h \, f(t) \, (S^z_1-S^z_2) \, ,  
\end{equation}
where $S^z_k = \sum_{i=1}^L S^z_{i,k}$, $f(t) = t/\tau$ for $0 < t \leq \tau$,
and $f(t)= 2 - t/\tau$ for $\tau < t \leq 2\tau$. The full Hamiltonian is
$H_{\text{tot}}(t) = H + h(t)$. We choose the field strength $h=0.5$ for all
simulations and vary the sweep time $\tau$.

\begin{figure}[t]
\includegraphics[width=0.80\columnwidth]{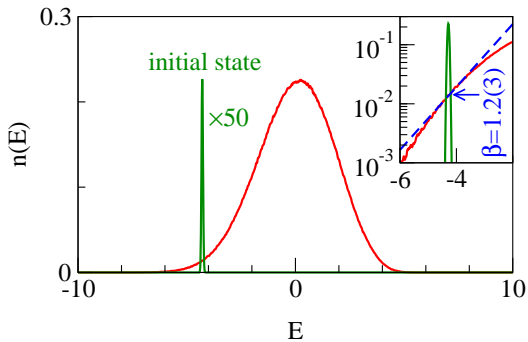}
\caption{DOS $n(E)$ for the Heisenberg $S=1/2$ ladder in Eq.\
(\ref{ham}) with $J_\perp = 0.2 \, J_\parallel$, $\Delta=0.6$, and $L=11$.
Due to the method used to obtain the numerical data, the energy resolution
$0.007(6)$ is high but finite \cite{SM}. The initial state prepared is
also indicated. Inset: The same as the main panel but in a semi-log plot
and with the inverse temperature $\beta = 1.2(3)$ indicated.}
\label{fig2}
\end{figure}

To specify a quantity that plays the role of temperature, we must have
information on the DOS of $H$. Since the
numerical diagonalization of $H$ is unfeasible for the system sizes we are
interested in, we resort to the numerical method described in \cite{hams2000}. This
method is incapable of resolving individual energy eigenvalues but captures
rather accurately the coarser features of the DOS \cite{SM}. The result for
our Hamiltonian is displayed in Fig.\ \ref{fig2}.

We choose the initial energy $E_\text{ini}$ to locate the process at a
non-peculiar temperature regime, i.e., neither extremely high nor very
low (nor negative) temperatures but an intermediate regime on the natural
scale of the model: $\beta \sim 1/J_\parallel$. To this end, we prepare an
initial state that is energetically firmly concentrated at $E_\text{ini}
= -4.2(8)$ for $L=11$. Using the definition $\beta = \text{d} / \text{d} E \ln
n(E)$ yields $\beta = 1.2(3)$. It is worth pointing out that, in this
energy regime, $\beta$ does not vary much on an interval of ca.\ $2
J_\parallel$, which is about the overall scale of the work required
for our process.

{\it Preparation and characterization of the initial state.} We prepare
a state of the form 
\begin{equation}
\left| \Psi(a,E_\text{ini}) \right> = \frac{e^{-a(H-E_\text{ini})^2/4} \left|
\Phi \right>} {\left< \Phi \right| e^{-a(H-E_\text{ini})^2/2} \left| \Phi
\right>} \, ,
\label{inistat}
\end{equation}
where $| \Phi \rangle$ is a random state drawn according to the Haar measure
on the total Hilbert space. Obviously, $\left | \Psi(a,E_\text{ini}) \right>$
is always centered at the energy $E_\text{ini}$ with a variance $\propto 1/a$.
Clearly, since $| \Phi \rangle$ is random, any quantity calculated from
$\left| \Psi(a, E_\text{ini}) \right>$ is random. However, as shown (and applied
\cite{steinigeweg2014-1, steinigeweg2014-2, steinigeweg2016}) in the context of
typicality, the average $\bar{Q}$ of any quantity $Q$ calculated from $\left|
\Psi(a, E_\text{ini}) \right>$ equals the $Q$ calculated from the mixed state
$\rho \propto e^{-a(H-E_\text{ini})^2 /2}$. Moreover, the ``error'' $\epsilon^2
= \bar{Q^2} - \bar{Q}^2$ scales as $\epsilon \propto \text{Tr} \{ e^{-a(H -
E_\text{ini})^2/2}) \}^{-1/2}$ and is very small if the Hilbert space is large
(but $a$ is not too large). Investing with reasonable computational effort, we are
able to reach $a = 1000$. In this regime, $\epsilon$ is negligibly small
\cite{SM}.

Using the same method as for calculating the DOS, we visualize the
probability distribution of a state $\left| \Psi(a,E_\text{ini}) \right>$ in Fig.\
\ref{fig2}. Clearly, this distribution is firmly concentrated at $E_\text{ini}
= -4.2(8)$.

\begin{figure}[t]
\includegraphics[width=0.80\columnwidth]{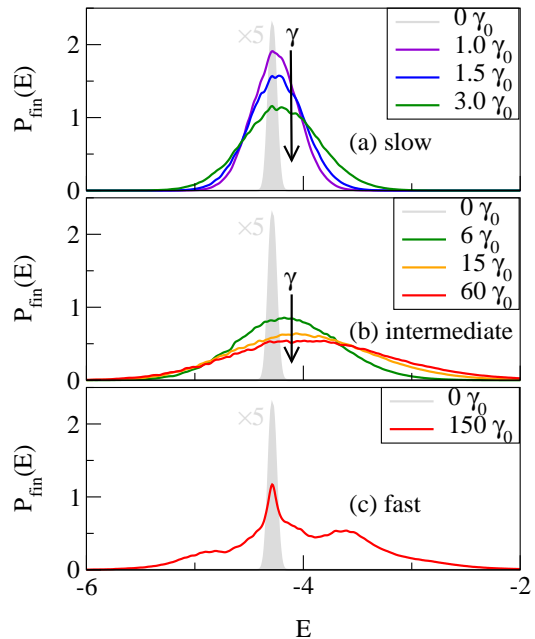}
\caption{Probability distribution $P_{\text{fin}}(E)$ of the final state
for (a) weak, (b) intermediate, and (c) fast driving. (Remaining parameters
are identical to Fig.\ \ref{fig2}.) Due to the initial
state being almost an energy eigenstate, this distribution almost
coincides with the probability distribution of work.}
\label{fig3}
\end{figure}

{\it Process, final energy distribution, and JR.} Now, we perform the
simulation of the actual process. To this end, we propagate  $\left|
{\Psi(a,E_\text{ini})} \right>$ in time according to the Schr\"odinger
equation using the time-dependent Hamiltonian $H_\text{tot}(t) = H + h(t)$
\cite{SM}. We do so for different sweep rates $\gamma = 1/(2 \tau)$, ranging
from slow driving $\gamma_0 = 2.6 \cdot 10^{-4}$ to fast driving at $\gamma
= 150 \gamma_0$. This yields a set of final energy-probability distributions
$P_{\text{fin}}(E, \gamma)$, see Fig.\ \ref{fig3}. Clearly, these
distributions shift towards higher energies and broaden with increasing
$\gamma$. Furthermore, they develop distinctly non-Gaussian features.

\begin{figure}[t]
\includegraphics[width=0.95\columnwidth]{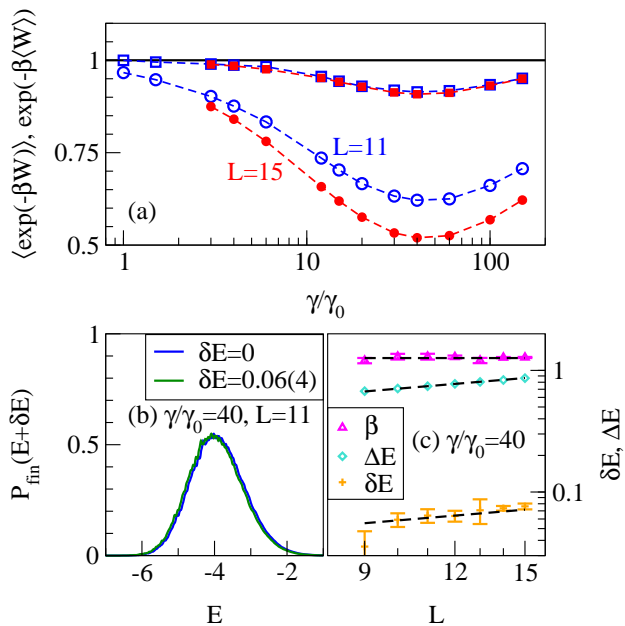}
\caption{(a) Averages $\langle e^{-\beta W} \rangle$ (squares) 
and $e^{-\beta \langle W 
\rangle}$ (circles) as a function of the process rate $\gamma$ for two different $L=11,
15$ and initial energies corresponding to the inverse temperature $\beta
\approx 1.2$. (b) Final distribution $P_\text{fin}(E)$ and fictitious
distribution $P_\text{fin}(E + \delta E)$ for $L=11$ and the rate $\gamma = 40
\, \gamma_0$, where the deviation of $\langle e^{-\beta W} \rangle$ from $1$ is
largest in (a). (c) Finite-size scaling of $\beta$, $\Delta E$, and $\delta E$
for the $\gamma$ in (b). All error bars indicated in (b) correspond to errors
resulting when determining $\beta$ by fitting locally the DOS.}
\label{fig4}
\end{figure}

Let us compare this result against the JR which here, since initial and final
Hamiltonian are the same, reads
\begin{equation}
\langle e^{-\beta W} \rangle =  \int P_W(W) \, e^{-\beta W} \, \text{d}W = 1
\, . \label{jar}
\end{equation}
If the initial state was a true energy eigenstate at energy $E =
E_\text{ini}$, then it would be justified to infer the actual probability
distribution of work $P_W$ from $P_{\text{fin}}$ as $P_W(W) = P_{\text{fin}}(W +
E_\text{ini})$. In this case the latter expression could be used to check
Eq.\ (\ref{jar}) directly. Given the ``narrowness'' of $P_{\text{ini}}$,
it seems plausible that the actual work-probability distribution $P_W(W)$
must be close to $P_{\text{fin}}(W + E_\text{ini})$. However, since
$P_{\text{ini}}$ is not precisely a $\delta$-distribution, one cannot,
strictly speaking, conclude from $P_{\text{fin}}$ onto $P_W$. To nonetheless
do so, we employ a further assumption, namely, that the true $P_W$ computed
from an actual initial $\delta$-function would not change much under variation
of the position of the initial $\delta$-peak on the order of the width of
$P_{\text{ini}}$, i.e., $\approx 0.25 J_\parallel$. Under this assumption,
the l.h.s.\ of  Eq.\ (\ref{jar}) may be cast into the form \cite{SM}
\begin{equation}
\frac{\int P_{\text{fin}}(E_{\text{fin}}) \, e^{-\beta E_{\text{fin}}} \,
\text{d}E_{\text{fin}}}{\int P_{\text{ini}}(E_{\text{ini}}) \, e^{-\beta
E_{\text{ini}}} \, \text{d}E_{\text{ini}}} = \int P_W(W) \, e^{-\beta W} \,
\text{d}W \, . \label{trick}
\end{equation}
Thus, the l.h.s.\ of Eq.\ (\ref{trick}) yields $\langle e^{-\beta W} \rangle$
based on $P_{\text{fin}}, P_{\text{ini}}$ for different sweep rates $\gamma$.
The closeness of the outcome to $1$ indicates how well the JR is fulfilled.
As shown in Fig.\ \ref{fig4}, for very slow processes the l.h.s.\ of Eq.\
(\ref{trick}) is practically $1$, while there is a difference of ca.\ $0.1$ for
faster processes. Comparing this difference to the deviation of $
e^{-\beta\langle W \rangle}$ from $1$, points to the approximate validity rather
than a strong violation of the JR. 

To judge further on this, we use the following scheme: For every actual
$P_{\text{fin}} (E_{\text{fin}})$ there is a fictitious probability
distribution ${\cal P}(E_{\text{fin}}) := P_{\text{fin}}(E_{\text{fin}}
+\delta E)$, which is identical in shape but shifted in energy by $\delta
E$ and fulfills the JR exactly, i.e., Eq.~(\ref{trick}) with $P_{\text{fin}} 
\rightarrow{\cal P}$ is identical to 1. We use this equation
to identify $\delta E$ and get $\delta E = 0.06(4)$ for $L=11$ and for $\gamma
= 40 \gamma_0$ where the deviation of the l.h.s.\ of Eq.\ (\ref{trick}) from
$1$ is largest. In Fig.\ \ref{fig4} (b) we display ${\cal P}(E_{\text{fin}})$
together with $P_\text{fin}(E_{\text{fin}})$. Clearly, the difference is hardly
visible as $\delta E$ is much smaller than the standard deviation $\Delta E =
0.7(4)$ of either distribution. Thus, while the JR is clearly violated, the
smallness of $\delta E$ indicates that this violation is remarkably small.

It should be stressed that the JR exponentially amplifies errors in the
negative tail of the distribution \cite{halpern2016}, i.e., a tiny lack of
statistics in this tail can in principle result in a large deviation from
the JR (the observed non-negativity of $\delta E$ probably reflects this
occurrence). This implies that one needs roughly an exponential number of
samples to get a good estimate of the average exponentiated work
\cite{jarzynski2006} even for initial canonical states. In this sense the
observed deviation from the JR of at most $10\%$, obtained from a single
wave function, is indeed small and a central result of
our Letter.

For all other sweep rates, $\delta E$ turns out to be even smaller, thus
rendering the actual work-probability distribution even closer to the
fictitious one. Note that the fictitious probability distribution
${\cal P}(E_\text{fin})$ introduced is certainly not the only choice
possible. However, it allows for a very natural interpretation.
 
{\it Finite-size scaling.} Finally, we perform a finite-size scaling for
$L=9, \ldots, 15$ and $E_\text{ini} = -0.42 (L-1)$, yielding $\beta \approx
1.2$ within the attainable precision. Focusing on $\gamma = 40 \gamma_0$
with the largest violation of the JR, we depict the scaling $\Delta E (L)$
and $\delta E (L)$ in Fig.\ \ref{fig4} (c). While $\Delta E (L)$ follows
a ``trivial'' upscaling $\Delta E (L) \propto \sqrt{L}$ \cite{SM}, giving a precise
statement on $\delta E (L)$ remains challenging. Given the error bars shown,
resulting from errors when determining $\beta$ by fitting \cite{SM}, a very
reasonable guess is $\delta E (L) \propto \sqrt{L}$ and indicated in Fig.\
\ref{fig4} (c). Then, $\delta E (L)/\Delta E (L) = \text{const}.$ and we
can expect that the JR remains valid to very good approximation for $L \to
\infty$. This is another central result of our work.

{\it Conclusions.} In summary, we have studied the validity of the JR
for non-canonical initial states that are pure states close to energy
eigenstates. To this end, we have performed large-scale numerics to
first prepare typical states of such kind and then to propagate these
states under a time-dependent protocol in a complex quantum system of
condensed-matter type. While we have found violations of the JR in our
non-equilibrium scenario, we have demonstrated that these violations are
remarkably small and point to the approximative validity of the JR in
a moderately sized system already. Furthermore, our systematic
finite-size analysis has not shown indications that this result changes
in the thermodynamic limit of very large systems.

While this result cannot be simply explained by the equivalence of
ensembles, it indicates the validity of the ETH for a non-trivial
operator being the ``operator of exponentiated work''. This validity
is surprising due to the structure of this operator but also since
the ETH is commonly associated with equilibrium properties, while
the JR addresses non-equilibrium processes. Promising directions of
future research include the generality of our findings for a wider
class of systems and protocols, the necessity of the two-measurement
scheme, as well as the dependence of the work distribution as such
on the type of initial condition realized.

{\it Acknowledgements.} The authors gratefully acknowledge the computing 
time granted by the JARA-HPC Vergabegremium and provided on the JARA-HPC 
Partition part of the supercomputer JUQUEEN~\cite{JUQUEEN15} at Forschungszentrum J\"ulich.
We are thankful for valuable insights
and fruitful discussions at working group meetings of the
COST action MP1209.

\section{Supplemental material}
\subsection{Time-dependent Schr{\"o}dinger equation}

The full Hamiltonian of the spin-$1/2$ ladder system reads ${\cal H}(t) = H + h(t)$,
where $H$ and $h(t)$ are defined by Eqs.\ (4) and (5) in the main text. Here, to shorten
notation, we write ${\cal H}(t)$ instead of $H_\text{tot}(t)$. The time evolution of
the system is governed by the time-dependent Schr{\" o}dinger equation (TDSE) (in units
of $\hbar =1$)
\begin{equation}
i\frac{\partial}{\partial t} |\Psi(t)\rangle = {\cal H}(t) |\Psi(t)\rangle \, ,
\end{equation}
where $|\Psi(t)\rangle$ is the wave function of the system. 
The solution of the TDSE can be written as
\begin{eqnarray}
|\Psi(t+\delta t)\rangle &=& U(t+\delta t,t) \, | \Psi(t)\rangle \cr
&=& \exp_+ \left( -i \int_t^{t +
\delta t} {\cal H}(u) \, \text{d}u \right ) \, |\Psi(t)\rangle,
\end{eqnarray}
where $\delta t$ is the time step. For small $\delta t$, the Hamiltonian is considered to
be fixed in the time interval $[t, t+\delta t]$ and then the time-evolution operator may be
approximated by
\begin{equation}
U(\delta t) = U(t+\delta t,t) = \exp(-i {\cal H}(t + \delta t/2) \delta t) \, .
\end{equation}

We solve the TDSE using a second-order product-formula algorithm \cite{DeRaedt87, DeRaedt06}.
The basic idea of the algorithm is to use a second-oder approximation of the time-evolution
operator $U(\delta t)$, given by
\begin{equation}
\widetilde{U}_2(\delta t) = e^{-i\delta t {\cal H}_k/2} \cdots e^{-i\delta t {\cal H}_1/2}
e^{-i\delta t {\cal H}_1/2} \cdots e^{-i\delta t {\cal H}_k/2} \, ,
\end{equation}
where ${\cal H}={\cal H}_1 + \cdots +{\cal H}_k$. The approximation is bounded by 
\begin{equation}
|| U(\delta t) - \widetilde{U}_2(\delta t) || \ll c_2 \, \delta t^3 \, ,
\end{equation}
where $c_2$ is a positive constant.

In practice, we use an XYZ decomposition for the Hamiltonian according to the $x$, $y$, and
$z$ components of the spin operators, i.e., ${\cal H}={\cal H}_x +{\cal H}_y +{\cal H}_z$.
The computational basis states are eigenstates of the $S^z$ operators. Thus, in this
representation $e^{-i \delta t {\cal H}_z}$ is diagonal by construction, and it only changes
the input state by altering the phase of each of the basis vectors. By an efficient basis
rotation into the eigenstates of the $S^x$ or $S^y$ operators, the operators $e^{-i \delta t
{\cal H}_x}$ and $e^{-i \delta t {\cal H}_y}$ act as $e^{-i \delta t {\cal H}_z}$.

\subsection{Initial state}

The initial state is obtained by a Gaussian projection of a random state drawn at random
according to the Haar measure on the total Hilbert space of the system,
\begin{equation}
|\Psi(a,E)\rangle = \frac{e^{-a(H-E)^2/4} \, |\Phi\rangle}{\langle \Phi| \, e^{-a(H-E)^2/2}
\, |\Phi\rangle} \, , \label{inistate}
\end{equation}
where $1/a$ characterizes the variance of the Gaussian projection and $H$ is the Hamiltonian
at $t=0$. This calculation is performed by employing the Chebyshev-polynomial representation
of a Gaussian function, properly generalized to matrix-valued functions \cite{TalEzer84,
Dobrovitski03}, and yields numerical results which are accurate to about $14$ digits.

In general, a function $f(x)$ whose values are in the range $[-1,1]$ can be expressed as
\begin{equation}
f(x) = \frac{1}{2} c_0 T_0(x)+\sum_{k=1}^{\infty} c_k T_k(x) \, ,
\end{equation}
where $T_k(x) = \cos(k \arccos x)$ are Chebyshev polynomials and the coefficients $c_k$ are
given by
\begin{equation}
c_k = \frac{2}{\pi} \int_{-1}^1 \frac{\text{d}x}{\sqrt{1-x^2}} \, f(x) \, T_k(x) \, .
\end{equation}
Let $x=\cos \theta$, then $T_k(x) = \cos (k \theta)$ and
\begin{eqnarray}
c_k&=&\frac{2}{\pi}\int_0^\pi f(\cos \theta) \cos (k \theta) \, \text{d} \theta   \cr
&=& \text{Re} \left [\frac{2}{N}\sum_{n=0}^{N-1} f \left(\cos\frac{2\pi n}{N} \right)
e^{2\pi i nk/N} \right] \, ,
\label{fftck}
\end{eqnarray}
which can be calculated by the fast Fourier transform (FFT).

For the operators $f(H) = e^{-a(H-E)^2/4}$, we normalize $H$ such that $\widetilde{H} = 
H/||H||$ has eigenvalues in the range $[-1,1]$ and put $\widetilde{a} = a ||H||$ and
$\widetilde{E}=E/||H||$. Then
\begin{equation}
f(\widetilde{H}) = e^{-\widetilde{a}(\widetilde{H}-\widetilde{E})^2/4} = \sum_{k=0}^{\infty}
c_k T_k(\widetilde{H}) \, ,
\label{cheby}
\end{equation}
where $\{c_k\}$ are the Chebyshev-expansion coefficients calculated from Eq.\ (\ref{fftck})
and the Chebyshev polynomial $T_k(\widetilde{H})$ can be obtained by the recursion relation
\begin{equation}
T_{k+1}(\widetilde{H}) - 2 \widetilde{H}T_{k}(\widetilde{H}) + T_{k-1}(\widetilde{H}) = 0
\end{equation}
with $T_0(\widetilde{H})=1$ and $T_1(\widetilde{H})=\widetilde{H}$.

In practice, the coefficients $c_k$ become exactly zero for a certain $k \geq K$. Hence, we
have an exact representation of the Gaussian projection up to a sum of $K$ terms in Eq.\
(\ref{cheby}). Note that the Chebyshev algorithm can only be efficiently applied to solve
the TDSE if the total Hamiltonian is time-independent \cite{Dobrovitski03}.

\subsection{Density of states}
\label{dos}

The density of states (DOS) of a quantum system may, on the basis of a time evolution, 
be defined as 
\begin{equation}
n(E)=\sum_n \delta(E-E_n)=\frac{1}{2\pi}\int_{-\infty}^{+\infty} e^{itE} \, \text{Tr} \{
e^{-itH} \} \, \text{d}t \, ,
\end{equation}
where $H$ is the Hamiltonian of the system at $t=0$ and $n$ runs over all eigenvalues $E_n$
of $H$. The trace in the integral can be estimated from the expectation value with respect
to a random vector, \textsl{e.g.}, by exploiting quantum typicality \cite{Hams00,
Steinigeweg14}. Thus, we have
\begin{equation}
\frac{\text{Tr} \{ e^{-itH} \}}{D} \approx \langle \Phi(0)|e^{-itH} |\Phi(0)\rangle =
\langle \Phi(0)|\Phi(t)\rangle \, ,
\end{equation}
where $D$ is the dimension of the Hilbert space and $|\Phi(0)\rangle$ is a pure state drawn
at random according to the unitary invariant measure (Haar measure) and the error scales
with $1/\sqrt{D}$. $|\Phi(t) \rangle = e^{-itH} |\Phi(0) \rangle$ can be efficiently computed
by the second-oder product-formula algorithm. Therefore, the DOS can be conveniently calculated
by FFT,
\begin{equation}
n(E)\approx C \int_{-\Theta}^{+\Theta} e^{itE} \langle \Phi(0)|\Phi(t)\rangle \text{d}t \, , 
\end{equation}
where $C$ is a normalization constant and $\Theta$ is the time up to which one has to integrate 
the TDSE in order to reach the desired energy resolution $\pi/\Theta$. The Nyquist sampling
theorem gives an upper bound to the time-step that can be used. For the systems considered in
the present paper, this bound is sufficiently small to guarantee that the errors on the
eigenvalues are small, see Ref.~\cite{Hams00} for a derivation of bounds, etc.

Similarly, we can obtain the local DOS (LDOS) of the system for a particular pure state
$|\Psi\rangle$, such as the initial state and final state of the system, by the formula
\begin{eqnarray}
P(E)&=&\sum_n d_n^2 \, \delta(E-E_n) = \sum_n |\langle E_n|\Psi\rangle|^2 \, \delta(E-E_n) \cr
&=& \frac{1}{2\pi} \int_{-\infty}^{+\infty} e^{itE} \, \langle \Psi| \, e^{-itH} \, |\Psi
\rangle \, \text{d}t \\
&\approx& C \int_{-\Theta}^{+\Theta} e^{itE} \, \langle \Psi| \, e^{-itH} \, |\Psi
\rangle \, \text{d}t \, ,
\end{eqnarray}
where $|E_n\rangle$ are energy eigenstates and $d_n=\langle E_n|\Psi\rangle$. 
Note that the concept of typicality is not involved in the calculation of $P(E)$.

\subsection{Numerical simulation}

\begin{figure}[t]
\includegraphics[width=8cm]{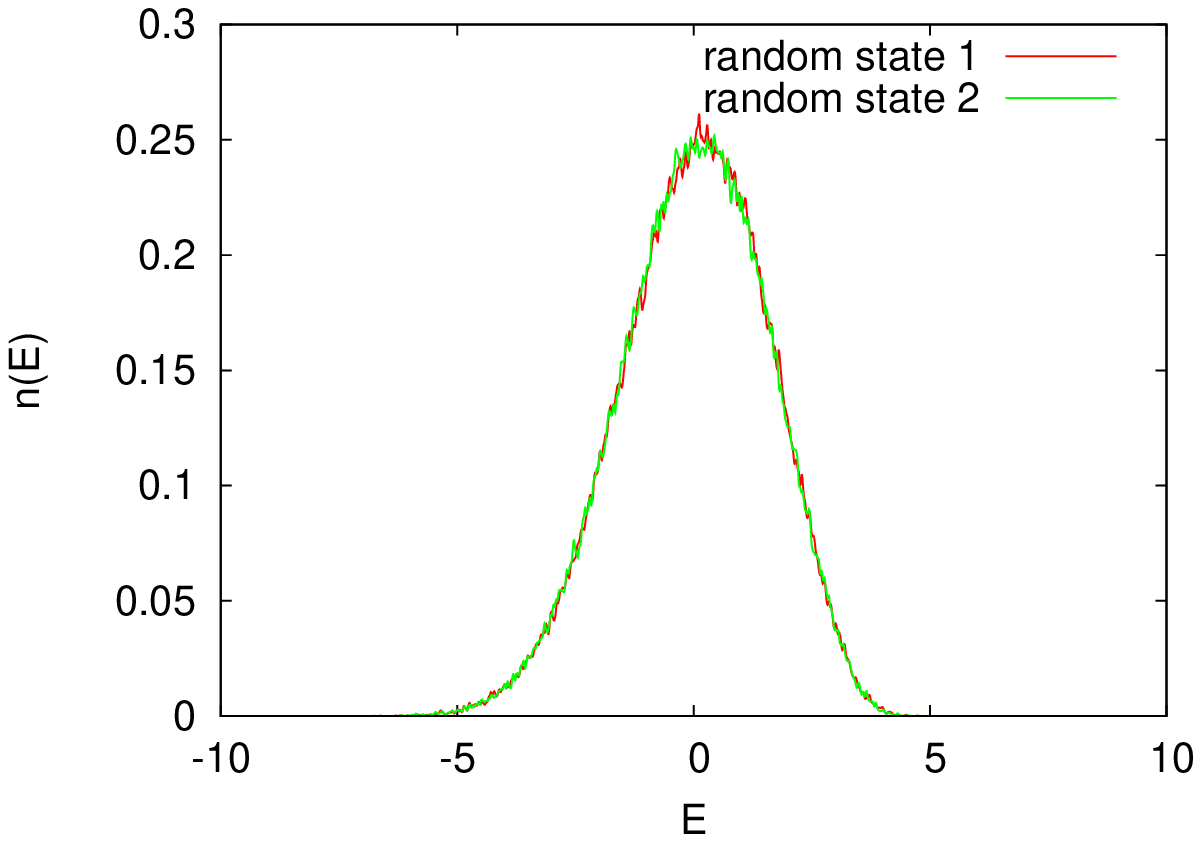}
\includegraphics[width=8cm]{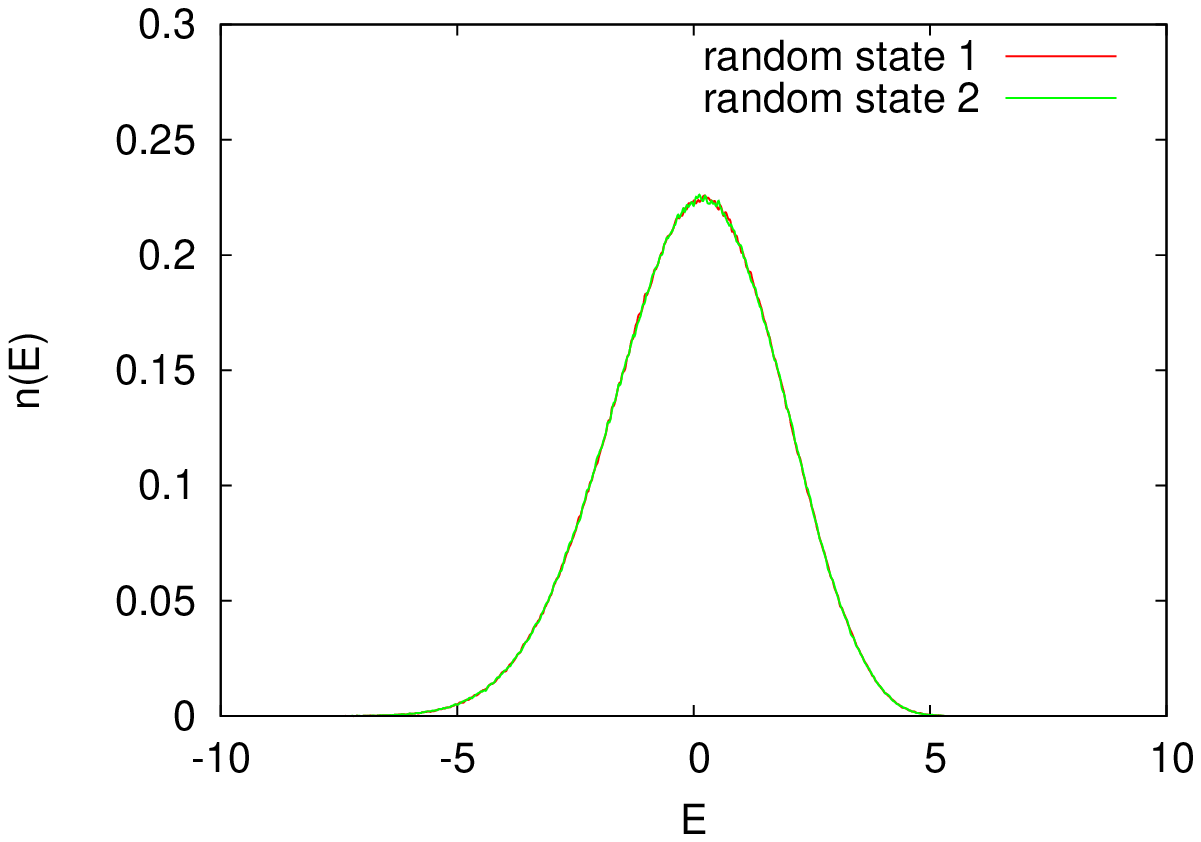}
\caption{Simulation results for the density of states $n(E)$ for a quantum
ladder system with $18$ spins (top) and $22$ spins (bottom). Two different random states
are used to compute $n(E)$ according to the algorithm described in the text. The system
Hamiltonian is defined by Eq.\ (4) in the main text.}
\label{figs1}
\end{figure}

The algorithm to compute the DOS $n(E)$ consists of the following steps:
\begin{enumerate}
	\item Generate a random state $|\Phi(0)\rangle$ at $t=0$.
	\item Copy this state to $|\Phi(t)\rangle$.
	\item Calculate $\langle\Phi(0)|\Phi(t)\rangle$ and store the result.
	\item Solve the TDSE for a small time step $\delta t$, replacing $|\Phi(t)\rangle$
	      by $|\Phi(t+\delta t)\rangle$. 
	\item Repeat $K$ times step $3$ and $4$. 
	\item Perform a Fourier transform on the tabulated result.
\end{enumerate}
In the simulation, we use $\delta t=0.02$ for the second-order product-formula algorithm
and repeat $K=4096 \times 5$ steps. The total simulation time is $\Theta=409.6$. Hence,
the energy resolution is about $\pi/\Theta\approx0.0077$. In principle, $n(E)$ may be
averaged over different random states. It turns out that this is only necessary for small
system sizes as the error scales with the square root of the dimension of the Hilbert space.
Figure~\ref{figs1} shows the simulation results for $n(E)$ obtained from two different
random states for systems with $18$ and $22$ spins. It can be clearly seen that the curves
$n(E)$ obtained for two different random states coincide apart from some small fluctuations. 
These fluctuations disappear for larger system sizes.

The strategy for numerically testing the Jarzynski relation is:
\begin{enumerate}
	\item Generate the initial state $|\Psi(a,E_{\text{ini}},t=0)\rangle$ by the
	      Chebyshev polynomial algorithm. 
	\item Calculate the LDOS $P_{\text{ini}}(E)$ for the initial state
	      $|\Psi(a,E_{\text{ini}},t=0)\rangle$.
	\item Solve the TDSE for the time-dependent Hamiltonian $H+h(t)$. 
	\item Calculate the LDOS $P_{\text{fin}}(E)$ for the final state
	      $|\Psi(a,E_{\text{ini}}, t=2\tau) \rangle $.
	\item Repeat from step $3$ for different process rates $\gamma = 1/2 \tau$.
\end{enumerate}
In the simulation, the parameters for the initial states are $a=1000$ and $E_{\text{ini}} 
= 0.42(L-1)$, $L = N/2$, where $N$ ranges from $18$ to $30$. We use $\delta t=0.02$ for the
second-order product-formula algorithm to solve the TDSE. After the whole process, we collect
the data sets of $n(E)$, $\langle E\rangle_{\text{ini}}$, $P_{\text{ini}}(E)$, $\langle E
\rangle_{\text{fin}}$, and $P_{\text{fin}}(E)$ for further analysis.

Our goal is to test the validity of the Jarzynski relation beyond the Gibbsian initial
state in isolated systems, i.e.,
\begin{equation}
\langle e^{-\beta W} \rangle = \int P_{\text{W}} (W) \, e^{-\beta W} \, \text{d}W
= e^{-\beta \Delta F} \, , \label{JR}
\end{equation}
where the work $W$ is defined as $W=E_{\text{fin}}-E_{\text{ini}}$ according to the
two-measurement scheme, $P_\text{W}(W)$ is the work probability, and $\Delta F$ is
the difference between the free energies of the two equilibrium states of the initial
and final Hamiltonian. Obviously, we need to calculate both sides of Eq.\ (\ref{JR}).
The right-hand side equals $1$ as the protocol we uses (see Eq.\ (5) in the main text)
ends with the same Hamiltonian as the initial one and hence $\Delta F=0$. Therefore,
we only need to calculate the left-hand side, which requires the information about the
inverse temperature $\beta$ and the work probability $P_\text{W}(W)$.

\subsubsection{Estimation of the inverse temperature $\beta$}

The initial state is narrowly centered at the initial energy $E_{\text{ini}}$ 
(as the standard deviation of $P_{\text{ini}}(E)$ is about $0.03$ for $a=1000$).
A microcanonical temperature can be calculated according to the standard formula 
\begin{equation}
\beta = \frac{\text{d} S}{\text{d} E} \, ,
\end{equation}
where $S = \ln n(E)$ is the microcanonical entropy.

We get $\beta$ from fitting $n(E)$ in the interval $[E_\text{ini}-\epsilon,E_\text{ini}
+ \epsilon]$, where $E_{\text{ini}}$ is the initial mean energy and $\epsilon$ is a
parameter to determine the range for fitting. The inverse temperature $\beta$ does not
vary significantly for $\epsilon \lesssim 0.5$ (see below).

\subsubsection{Estimation of the work probability $P_\text{W}(W)$}

As mentioned in the main text, we have two non-trivial distributions of
energy, a final and an initial one, from which $P_W$ must be inferred. Without
further assumptions, the attribution of a final and an initial distribution of
energy to one distribution of work cannot be entirely unique. Thus, we rely on
a further assumption: The probability distribution of work as arising from an
initial $\delta$ distribution w.r.t.\ energy could in principle vary strongly
with the position $E_\text{ini}$ at which this initial distribution is peaked.
Assuming that this is not the case on the small regime where $P_{\text{ini}}(E)$
takes on non-negligible values, the probability distributions of work and
energy are related as
\begin{equation}
\label{conv}  
P_{\text{fin}}(E_{\text{fin}}) = \int P_{\text{ini}}(E_{\text{ini}}) \,
P_W(E_{\text{fin}} - E_{\text{ini}}) \, \text{d}E_{\text{ini}} \, .
\end{equation}
Multiplying this equation by $e^{-\beta}E_{\text{fin}}$ and integrating over
$E_{\text{fin}}$, followed by a change of variables $E_{\text{fin}} \rightarrow
W + E_{\text{ini}}$ on the r.h.s.\ yields 
\begin{equation}
\frac{\int P_{\text{fin}}(E_{\text{fin}}) \, e^{-\beta E_{\text{fin}}} \,
\text{d}E_{\text{fin}}}{\int P_{\text{ini}}(E_{\text{ini}}) \, e^{-\beta
E_{\text{ini}}} \, \text{d}E_{\text{ini}}} = \int P_\text{W}(W) \, e^{-\beta W} \,
\text{d}W \, . \label{trick1}
\end{equation}
Thus, the l.h.s.\ of Eq.\ (\ref{trick1}) yields $\langle e^{-\beta W} \rangle$
based on $P_{\text{fin}}, P_{\text{ini}}$ for different sweep rates $\gamma$.

\subsubsection{Estimations of $\langle e^{-\beta W} \rangle$ and $e^{-\beta \langle W \rangle}$}

\begin{figure}[t]
\includegraphics[width=8cm]{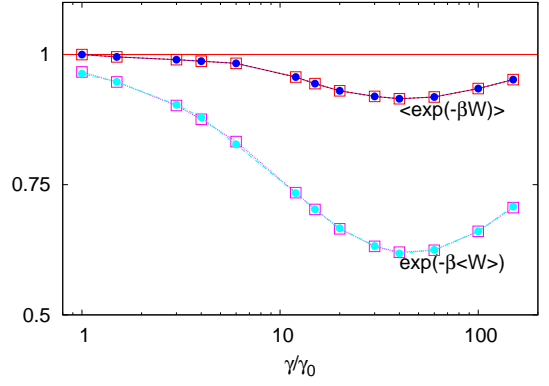}
\caption{Simulation results for $\langle e^{-\beta W} \rangle$ and $e^{-\beta \langle W
\rangle}$ as a function of the process rate $\gamma$ (normalized to the slowest rate
$\gamma_0=2.6\times 10^{-4}$) for two different random states $|\Phi\rangle$ (represented
by squares and circles) used to determine the initial states (see Eq.~(\ref{inistate})).
The system is a quantum ladder with $22$ spins. The initial energy corresponds to the
inverse temperature $\beta=1.23$ (see Fig.\ 2 (inset) in the main text).}
\label{figs2}
\end{figure}

In order to calculate the left-hand side of Eq.\ (\ref{JR}), i.e., $\langle e^{-\beta W}
\rangle$, we do not have to really calculate the work probability. As explained before,
the left-hand side can be expressed as
\begin{equation}
\langle e^{-\beta W} \rangle = \frac{\int P_{\text{fin}}(E_{\text{fin}}) \,
e^{-\beta E_{\text{fin}}} \, \text{d}E_{\text{fin}}}{\int P_{\text{ini}}(E_{\text{ini}})
\, e^{-\beta E_{\text{ini}}} \, \text{d}E_{\text{ini}}} \, .
\end{equation}
Similarly, we have
\begin{eqnarray}
\langle W \rangle &=& \int P_{\text{fin}}(E_{\text{fin}}) \, {E_{\text{fin}}} \, \text{d}
E_{\text{fin}} - \int P_{\text{ini}}(E_{\text{ini}}) \, {E_{\text{ini}}} \, \text{d}
E_{\text{ini}} \cr
&=& \langle E_{\text{fin}}\rangle -\langle E_{\text{ini}} \rangle
\end{eqnarray}
and
\begin{equation}
e^{-\beta\langle W \rangle} = e^{-\beta(\langle E_{\text{fin}}\rangle -\langle E_{\text{ini}}
\rangle)} \, .
\end{equation}
Hence, the calculations of $\langle e^{-\beta W}\rangle$ and $e^{-\beta \langle W \rangle}$ 
solely depend on the data set obtained from the simulation.

Figure~\ref{figs2} presents the simulation results for $\langle e^{-\beta W}\rangle$ 
and $e^{-\beta \langle W \rangle}$ as a function of the process rate of
the magnetic field imposed on the two legs of the ladder for $22$ spins. 
The inverse temperature $\beta$ is set to $1.23$ (see Fig.\ 2 in the main text). 
Two different random states are used to prepare the initial state of the system.
It can be seen that $\langle e^{-\beta W}\rangle$ and $e^{-\beta \langle W \rangle}$ 
calculated for these two different initial states do not differ much. Hence, it
is sufficient to study the Jarzynski relation only for one particular initial state.

\subsubsection{Error estimation of $\langle e^{-\beta W} \rangle$}

The main error of our overall analysis is set neither by our numerical methods
(finite time step, maximum time) nor by the specific realization of the initial
state. Instead, the main error results when determining the inverse temperature
$\beta$ by fitting locally the density of states. By varying the fit range
$[E_\text{ini}-\epsilon, E_\text{ini}+\epsilon]$ from $\epsilon=0.25$ to
$\epsilon = 0.5$, we find that the value of $\beta$ can be determined with a
precision of $\approx 5 \, \%$, see the small error bars in Fig.\ 4 (c) of
the main text. This precision implies that, for the sweep rate $\gamma/
\gamma_0 = 40$, the quantity $\langle e^{-\beta W} \rangle$ has a corresponding
error of $\approx 2 \, \%$. This error is smaller than the symbol size used and
not indicated explicitly in Fig.\ 4 (a) of the main text. We can therefore
exclude that the deviation of this quantity from $1$ for such values of $\gamma$
is an artifact of our approach.

However, the corresponding error for the shift $\delta E$ is much larger
since
\begin{equation}
\delta E = -\frac{1}{\beta} \, \ln  \langle e^{-\beta W} \rangle
\end{equation}
essentially is the logarithm of a small number $< 1$. Consequently, the
corresponding error can be as large as $\approx 30 \, \%$, see the error bars
in Fig.\ 4 (c) of the main text. Such errors of $\delta E$ are particularly
relevant for the quality of the finite-size scaling $\delta E(L)$ and thus taken
into account in the conclusions.

\subsection{Finite-size scaling}

A central result of our Letter concerns the upscaling of the
system towards the limit $L \rightarrow \infty$. It is instructive to compare
to a set of $M$ disconnected small systems: The work-probability
distribution in this case is a $M$-fold convolution of the work-probability
distribution as resulting for each small system. Since mean values are
additive under convolution, one gets the corresponding shift scaling as
$\delta E (M) = M \delta E (1)$. The standard deviation of the work-probability
distribution, however, scales under convolution as $\Delta E (M) = \sqrt{M}
\Delta E (1)$. This implies that, for large $M$, $\delta E (M)$ becomes
inevitably larger than $\Delta E (M)$ and thus the resulting work-probability
distribution is far away from fulfilling the JR. Therefore, in the limit of many
disconnected small systems, the JR is strongly violated whenever $\delta E (1)$
is non-zero. This finding is clearly different to a long connected ladder, as
discussed in our Letter.

\nocite{niemeyer2014}
\nocite{gemmer2014}
\nocite{schmidtke2016}

\bibliography{references}

\begin{thebibliography}{40}
\bibitem [{not()}]{note}%
  \BibitemOpen
  \href@noop {} {}\bibinfo {note} {Here, we assume that the spectrum is
  non-degenerate.}\BibitemShut {Stop}%
\end{thebibliography}

\end{document}